\documentclass[%
reprint,
superscriptaddress,
nofootinbib,
nobibnotes,
 amsmath,amssymb,
 aps,
longbibliography
]{revtex4-2}
\usepackage[english]{babel}
\usepackage{graphicx}
\usepackage{dcolumn}
\usepackage{bm}

\usepackage[dvipsnames]{xcolor}
\usepackage[colorlinks=true,citecolor=LimeGreen,linkcolor=blue, urlcolor = cyan]{hyperref} 

\usepackage{color}
\usepackage[english]{babel}

\newcommand{\Tr}{\mathop{\rm Tr}\nolimits}

\begin{document}

\title{Odd viscosity and anomalous Hall effect in two dimensional systems with smooth disorder}

\author{D.S. Zohrabyan}
\affiliation{L. D. Landau Institute for Theoretical Physics, 142432 Chernogolovka, Russia}
\affiliation{Moscow Institute of Physics and Technology, 141701 Dolgoprudny, Russia}
\author{M.M. Glazov}
\affiliation{Ioffe Institute, 194021 St. Petersburg, Russia }%


\begin{abstract} 
A microscopic theory of odd viscosity in two-dimensional electron systems with smooth disorder and spin-orbit interaction is developed. It is shown that spin-orbit scattering in presence of spin polarization induced by magnetic field gives rise to an off-diagonal component of the viscosity tensor. Hydrodynamic equations for spin and electric currents are derived for electrons interacting with smooth disorder. The contribution of odd viscosity to the anomalous Hall effect is calculated.
\end{abstract}

\maketitle

\textbf{1. Introduction.} In recent years, significant attention of researchers has been drawn to transport phenomena and effects in ultra-clean two-dimensional electronic systems, where scattering processes on static defects and phonons are suppressed compared to electron-electron scattering, scattering on rough channel boundaries, and smooth disorder~\cite{Levitov:2016aa,Crossno:2016aa,Bandurin1055,Moll1061,Krishna-Kumar:2017wn,PhysRevLett.117.166601}. In such systems, both ballistic and hydrodynamic regimes of electron propagation can be realized; in the latter, dissipation processes are controlled by electronic viscosity~\cite{gurzhi63,Gurzhi_1968}. Phenomena where the differences between ultra-pure systems and ordinary (diffusive) ones manifest most clearly are of particular interest. Among these are the normal and anomalous Hall effects and similar other spin-dependent phenomena~\cite{PhysRevLett.118.226601,PhysRevB.96.195401,Doornenbal_2019,Berdyugin162,PhysRevB.100.125419,PhysRevB.104.085434,PhysRevB.106.245415,grigoryan2023anomalous,Glazov_2021b,PhysRevB.106.L041407,PhysRevB.106.L081113,PhysRevResearch.3.033075,PhysRevB.104.184414,PhysRevB.103.125106}.

Usually, the hydrodynamic regime of quasiparticle propagation is associated with a situation where interparticle collisions dominate over any other dissipative processes, allowing the quasiparticle distribution function to be represented as an equilibrium one (Fermi-Dirac function for electrons or Bose-Einstein for excitons), whose parameters (temperature, chemical potential, average -- hydrodynamic -- velocity, i.e., a shift of the distribution function in $\mathbf k$-space) are determined by external influences~\cite{gantmakher87,mantsevich2024viscoushydrodynamicsexcitonsvan}. However, a quasi-hydrodynamic transport regime can be realized if quasiparticles interact with smooth disorder, see, e.g., ~\cite{PhysRevB.106.245415} and references therein. In this case, the relaxation times of the angular harmonics of the distribution function rapidly  decrease with increasing harmonic number, and the relaxation of the second angular harmonic, which determines viscosity, occurs four times faster than that of the first.

In connection with studies of hydrodynamic and quasi-hydrodynamic regimes of electronic transport, particular attention has been drawn to systems where there is an odd or off-diagonal component of the viscosity tensor, $\eta_{xy}=-\eta_{yx}$~\cite{Avron:1998aa,Banerjee:2017aa,PhysRevFluids.2.094101,fruchart2022odd,Fruchart:2023aa,g9yl-rtbr}. It is known that such a component of the viscosity tensor -- Hall viscosity -- arises due to the action of the Lorentz force on electrons in an external magnetic field~\cite{chapman1990mathematical,PhysRevLett.117.166601,PhysRevB.106.245415}. Contributions to $\eta_{xy} = -\eta_{yx}$ due to electron spin polarization and spin-orbit interaction are also possible in presence of spin polarization~\cite{PhysRevResearch.3.033075,grigoryan2023anomalous}. These contributions, as well as the Hall contribution, are caused by the violation of time-reversal symmetry. Our goal is to develop an analytical theory of odd viscosity caused by spin-orbit interaction and scattering on smooth disorder. It is shown that odd viscosity arises due to asymmetric scattering by smooth disorder. In contrast to previous studies, where odd viscosity is either due to the Lorentz force~\cite{PhysRevB.104.184414} or arises only due to the inhomogeneity of the electric field and spin polarization~\cite{PhysRevResearch.3.033075}, this contribution require neither interparticle collisions nor inhomogeneities of the electric field and spin polarization. A similar situation arises in the theory of the spin Hall effect in the hydrodynamic regime~\cite{PhysRevB.96.020401,Takahashi:2020um,10.1093/pnasnexus/pgae547,Huang:2025aa}, but to our knowledge, the contribution of scattering by smooth disorder to the odd ``spin'' viscosity has not been discussed before.

In this paper we derive macroscopic quasi-hydrodynamic equations for a spin-polarized electron gas interacting with smooth disorder in the presence of spin-orbit coupling, obtaining a microscopic expression for the odd viscosity $\eta_{xy}$. The anomalous Hall effect caused by odd viscosity is considered as an example of using a system of quasi-hydrodynamic equations.

\textbf{2. Model.}
Let us consider a two-dimensional degenerate electron gas in the $(xy)$ plane. An external electric field $\mathbf E$ is applied in the plane of electron motion $(xy)$, and an external magnetic field is perpendicular to it $\mathbf B \parallel \mathbf z$. Due to the Zeeman effect, the magnetic field creates a non-zero spin polarization of the electrons and also induces cyclotron motion of charge carriers under the action of the Lorentz force. In this work, we neglect quantum effects associated with the formation of Landau levels. The Hall effect arising in confined systems leads, as known, to the formation of a Hall field $\mathbf E_H$, perpendicular to the external $\mathbf E$ and $\mathbf B$ fields~\cite{Hall:1879aa,Hall:1881aa}. As known, the Hall field $\mathbf E_H = \mathbf E_H^n+\mathbf E_H^a$ has two contributions: a ``normal'' one ($\mathbf E_H^n$), due to the action of the Lorentz force on electrons, and an anomalous one ($\mathbf E_H^a$), which is proportional to the electron spin polarization degree and related to spin-orbit interaction~\cite{Adams:1959aa,gy61,abakumov72,nozieresAHE,Sinitsyn_2007,dugaev_AHE,RevModPhys.82.1539,Ado_2015,PhysRevLett.126.036801}. The latter will be of interest to us here.

We introduce the $2\times 2$ spin density matrix $\hat{\rho}_{\mathbf k} = f_{\mathbf k} \hat{I} + \mathbf s_{\mathbf k} \cdot \hat{\bm \sigma}$, where $\mathbf k$ is the electron quasi-wave vector, $f_{\mathbf k} = (1/2) \Tr \{\hat{\rho}_{\mathbf k}\}$ is the spin-averaged occupancy of state $\mathbf k$, $\mathbf s_{\mathbf k} = (1/2) \Tr\{\hat{\rho}_{\mathbf k}\hat{\mathbf \sigma}\}$ is the electron spin distribution function. Here $\hat{I}$ is the identity matrix (hereafter omitted for brevity), $\hat{\bm \sigma}=(\hat{\sigma}_x,\hat{\sigma}_y,\hat{\sigma}_z)$ is a pseudovector composed of the Pauli spin matrices. In this work, we neglect the wave-vector-odd splitting of conducting band states by spin, so the density matrix for $\mathbf B \parallel \mathbf z$ is diagonal and parametrized by two distribution functions $f_{\mathbf k}^\pm = f_{\mathbf k} \pm s_{z, \mathbf k}$. For what follows, this representation of the density matrix will be most convenient. The density matrix obeys the Boltzmann kinetic equation, from which one can obtain equations for the distribution functions $f_{\mathbf k}^{\pm} \equiv f_{\mathbf k}^{\pm}(\mathbf r, t)$ in the following form~\cite{grigoryan2023anomalous}:
\begin{equation}\label{Model:general form of kinetic equation}
\frac{\partial f_{\mathbf k}^{\pm}}{\partial t} + \mathbf v_{\mathbf k}\frac{\partial f_{\mathbf k}^{\pm}}{\partial \mathbf r} + \mathbf F \cdot \frac{1}{\hbar}\frac{\partial f_{\mathbf k}^{\pm}}{\partial \mathbf k} = Q^{\pm}_{dis}\{f_{\mathbf k}^{\pm}\}
\end{equation}
Here $\mathbf v_{\mathbf k} \equiv \hbar^{-1} \partial\varepsilon_{k}/\partial \mathbf k= \hbar \mathbf k/m$ is the "kinematic" electron velocity, $\varepsilon_{k} = \hbar^2 k^2/2m$ is its dispersion, and $m$ is the effective mass,
\begin{equation}
\mathbf F = e(\mathbf E + \mathbf E_H) + \frac{e}{c}[\mathbf v_{\mathbf k}\times \mathbf B],
\label{force:F}
\end{equation}
is the force acting on the electron from the total electric field $\mathbf E+\mathbf E_H$ arising in the channel due to charge accumulation at its edges and the magnetic field $\mathbf B$, $e<0$ is the electron charge, $c$ is the speed of light in vacuum. Here we are interested only in effects related to the odd viscosity due to scattering, so we do not take into account here contributions to the anomalous Hall effect due to Berry curvature of energy bands (anomalous electron velocity) and associated wave packet shift contributions (side jump)~\cite{2020arXiv200405091G,Glazov2020b,Glazov_2021b,grigoryan2023anomalous}. On the right side of~\eqref{Model:general form of kinetic equation}
$Q_{dis}\{f_{\mathbf k}\}$ is the collision integral describing electron scattering on disorder.

We will consider only those scattering processes that leave electrons in the conduction band, and we will also neglect contributions from spin-flip processes. The spin-orbit interaction is taken into account using $\mathbf{k\cdot p}$ model, it arises due to the valence band states admixture to the conduction band  \cite{2020arXiv200405091G}. The scattering matrix element from a state with wave vector $\mathbf k'$ to state $\mathbf k$ in the first non-vanishing approximation in spin-orbit interaction can be written as
\begin{equation}\label{Hamiltonian and Q: V_k'k}
V_{\mathbf{k'k}}^{\pm} = U_c(\mathbf{k-k'}) \pm i\xi U_v(\mathbf{k-k'})[\mathbf k'\times \mathbf k]_z, \end{equation}
where the superscripts $\pm$ enumerate the states with spin component  $s_z =\pm 1/2$, $\xi$ is a quantity characterizing the strength of spin-orbit interaction~\cite{2020arXiv200405091G,averkiev02}, $U_{c,v}(\mathbf q)$ are the Fourier transforms of the impurity potentials in the conduction and valence bands, respectively, which can differ if impurity potential is not Coulombic. A generalization for the multi-band model is given in Ref. \cite{2020arXiv200405091G}.
The dimensionless small parameter characterizing the strength of spin-orbit interaction is $\xi k_F^2 \ll 1$, we also assume that $\varepsilon_F \ll E_g$, where $\varepsilon_F$ and $k_F$ are the electron Fermi energy and Fermi wave vector, $E_g$ is the band gap.

The electron-impurity collision integral can be written in general form as \cite{Sturman_1984}
\begin{equation}\label{Hamiltonian and Q: Q_dis}
    {Q_{dis}^{\pm}\{f_{\mathbf k}\} = \sum_{\mathbf k'} (W_{\mathbf{kk'}}^{\pm}f_{\mathbf k'}^{\pm} - W_{\mathbf{k'k}}f_{\mathbf k}^{\pm})},\end{equation}
where
\begin{equation}\label{Hamiltonian and Q: W_k'k}
W_{\mathbf{k'k}}^{\pm} = \frac{2\pi}{\hbar}\delta(\varepsilon_{k} - \varepsilon_{k'})\{ t_0(\theta) \pm t_1(\theta)\sin\theta\}, \;\;\; \theta = \angle \mathbf{k'k}.
\end{equation}
Here the first term describes spin-independent scattering which appears in the Born approximation, and the second term describes asymmetric spin-dependent scattering (Mott effect)~\cite{dyakonov71a,abakumov72,PhysRevLett.103.186601,dyakonov_book},  that arises in third order over scattering matrix element \eqref{Hamiltonian and Q: V_k'k}, $t_0(\theta)$, $t_1(\theta)$ are even functions of angle $\theta$, with $t_0(\theta) = n_{imp}|U_c(\theta)|^2$, where $n_{imp}$ is the impurity concentration,
\begin{equation}\label{Hamiltonian and Q: t_1}
t_1(\theta) = 2\pi Dk_F^2\xi n_{imp}\cdot U_v(\theta)\int_0^{2\pi}\frac{d\varphi}{2\pi}U_c(\varphi)U_c(\theta - \varphi),
\end{equation}
where $D = m/2\pi\hbar^2$ is the density of states of per spin component, also we introduced the notation $U_{c,v}(\theta)\equiv U_{c,v}(2k|\sin\theta/2|)$. We stress that in the Born approximation, terms like $\hat{\mathbf \sigma} [\mathbf k \times \mathbf k']$ in the scattering amplitude contain an imaginary unity and do not interfere with the spin-independent part, so the asymmetric contribution \eqref{Hamiltonian and Q: t_1} does not arise in the Born approximation. It is because of Hermiticity of the Hamiltonian.
Let us present the nonequilibrium, i.e., induced by external electric and magnetic fields $\mathbf E$ and $\mathbf B$, correction to the distribution function in the form
\begin{equation}
\label{non:eq}
\delta f_{\mathbf k}^\pm = \sum_{n = 1}^\infty \left\{ f_{c,n}^\pm\cos n\varphi + f_{s,n}^\pm\sin n\varphi\right\},
\end{equation}
where $f_{c,n}^\pm$, $f_{s,n}^\pm$ are the electron energy-dependent expansion coefficients, $\varphi_{\mathbf k}=\angle \mathbf k,Ox$, and obtain
\begin{multline}\label{Hamiltonian and Q: Q with G_sk for pm}
Q_{dis}^\pm\{f_{\mathbf k}^\pm\} = -\sum_{n=1}^\infty \frac{f_{c,n}^\pm\cos n\varphi + f_{s,n}^\pm\sin n\varphi}{\tau_n} \\ \pm \sum_{n=1}^\infty\gamma_n(f_{s,n}^\pm\cos n\varphi - f_{c,n}^\pm\sin n\varphi).
\end{multline}
The sum in the first line of~\eqref{Hamiltonian and Q: Q with G_sk for pm} is the symmetric part of the collision integral, describing the relaxation of angular harmonics of the distribution function with relaxation times $\tau_n$ that are introduced in the standard way.
The contribution in the second line of~\eqref{Hamiltonian and Q: Q with G_sk for pm} corresponds to the asymmetric contribution to the collision integral, and the coefficients $\gamma_n$ describe the conversion of even and odd harmonics due to spin-orbit interaction:
\begin{equation}
\gamma_n = \frac{2\pi}{\hbar}D\int \frac{d\theta}{2\pi}t_1(\theta)\sin\theta \sin n\theta.
\end{equation}
The quantities $\tau_n$ and $\gamma_n$ depend on electron energy. Note that in the case of a short-range potential, only the first harmonic of the distribution function is present in the asymmetric part of the collision integral~\cite{Glazov_2021b, grigoryan2023anomalous,glazov_zphrabyan:ftt}, and the other contributions vanish: $\gamma_{n} = \gamma_1\delta_{1,n}$. In the case of a long-range potential, all coefficients are non-zero $\gamma_n \neq 0$, in particular, the coefficient at the second angular harmonic $\gamma_2$ determines the odd viscosity. %
Below, we consider the quasi-hydrodynamic regime, keeping only the first two harmonics in the collision integral \eqref{Hamiltonian and Q: Q with G_sk for pm}. Although there is no formal small parameter in the problem, $\tau_n\propto n^{-2}$ and higher harmonics of the distribution function are neglected because of their small numerical coefficients.

\textbf{3. Odd viscosity.} The components of the viscosity tensor for two-dimensional electrons can be derived following the standard approach outlined, e.g., in~\cite{ll10_eng}. Considering a locally equilibrium distribution of electrons with hydrodynamic velocity in each spin branch $\mathbf u^\pm\equiv \mathbf u^\pm(\mathbf r,t) = N_\pm^{-1}\sum_{\mathbf k} \mathbf v_{\mathbf k} f^\pm_{\mathbf k}$, smoothly (on the scale of the mean free path) depending on coordinates, where $N^\pm = \sum_{\mathbf k} f_{\mathbf k}^\pm$ are the concentrations of electrons with a given spin projection, one can represent the components of the momentum flux tensor as $\Pi_{\alpha\beta}^\pm = m\sum_{\mathbf k} v_{\alpha, \mathbf k} v_{\beta,\mathbf k} f_{\mathbf k}$. In the case of axial symmetry of interest to us, the term with the gradient of the momentum flux can be represented in the following convenient form
\begin{equation}\label{vis:def}
-\frac{\mathbf \nabla \hat{\Pi}^\pm}{m} = \Delta (\eta_{xx}^\pm N^\pm \mathbf u^\pm) + \Delta[\eta_{xy}^\pm N^\pm \mathbf u^\pm \times \hat{\mathbf z}],
\end{equation}
where $\hat{\mathbf z}$ is unit vector of z-axis, and two independent components of the (first) viscosity tensor are introduced: $\eta \equiv \eta_{xx}^\pm$ (diagonal viscosity) and $\eta_{xy}^\pm = -\eta_{yx}^\pm$, the off-diagonal (odd) or Hall viscosity~\cite{Avron:1998aa,Banerjee:2017aa,PhysRevFluids.2.094101,fruchart2022odd,Fruchart:2023aa}. The latter is the subject of our interest. Note that in our problem formulation, the second viscosity does not manifest itself (general analysis is given, e.g., in~\cite{PhysRevB.103.235305}).\footnote{In \eqref{vis:def}, unlike the standard representation of the Navier-Stokes equations~\cite{ll6_eng}, the terms with the product of viscosity and concentration are under the Laplace operator. This is related to the specifics of scattering on static disorder. For an incompressible electron liquid, this difference is not essential.}

\begin{figure}[t]
\centering
\includegraphics[width=\linewidth]{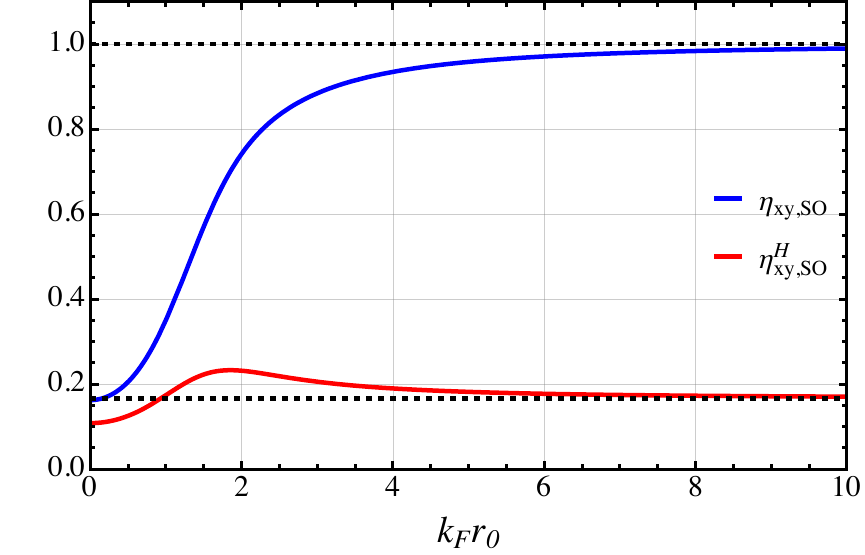}
\caption{The plot shows the dependence of odd viscosity $\eta_{xy,SO}$ (blue line) in the case of a Gaussian impurity potential \eqref{Odd viscosity: U_c,v gauss} on the correlation length  parameter $k_Fr_0$. The odd viscosity is normalized to its asymptotic value $\eta_{xy,SO}^{gauss}$ \eqref{Odd viscosity: eta_xy,SO^gauss} at $k_Fr_0 \rightarrow \infty$. The ``effective'' odd viscosity $\eta_{xy,SO}^H$ (red line), which enters the expression for the Hall field \eqref{Hall effect: E_H^a}, is also shown.}
\label{fig:2 eta_gauss}
\end{figure}

Expanding the distribution function in angular harmonics~\eqref{non:eq} and taking into account spatial gradients and magnetic field in the lowest (first) order, we obtain the diagonal viscosity
\begin{equation}
\label{eta:xx}
\eta_{xx}^\pm \equiv \eta^\pm = \frac{(v_F^\pm)^2 \tau_2(\varepsilon_F^\pm)}{4},
\end{equation}
where $v_F^\pm = \sqrt{2\varepsilon_F^\pm/m}$ is the electron velocity at the Fermi surface in the corresponding spin branch, and $\tau_2$, as indicated above, is the relaxation time of the second angular harmonic of the distribution function. In the relevant case of a small degree of electron polarization, the differences between $\eta^+$ and $\eta^-$ are negligible. The off-diagonal viscosity contains two contributions:
\begin{equation}
\eta_{xy}^{\pm} = \eta_{xy,H}^\pm + \eta_{xy,SO}^\pm,
\end{equation}
where
\begin{equation}
\label{Hall:eta}
\eta_{xy,H}^\pm = 2\omega_c\tau_{2}\cdot \eta
\end{equation}
is the Hall component of odd viscosity due to the action of the Lorentz force on electrons, where $\tau_2$ and $\eta$ are taken at the Fermi level, assuming magnetic field to be small $2\omega_c\tau_2\ll 1$~\cite{chapman1990mathematical,PhysRevLett.117.166601,PhysRevB.106.245415}; $\eta_{xy,SO}^\pm$ is the contribution to odd viscosity due to spin-orbit interaction~\cite{PhysRevResearch.3.033075,PhysRevB.96.020401,Takahashi:2020um,grigoryan2023anomalous}. The latter can be represented in a form similar to~\eqref{Hall:eta},
\begin{equation}
\label{so:eta}
\eta_{xy,SO}^\pm = \pm \gamma_2(\varepsilon_F^\pm)\tau_2(\varepsilon_F^\pm) \eta^\pm \equiv \pm \eta_{xy,SO}^s(\varepsilon_F^\pm).
\end{equation}
The odd viscosity of the electron gas due to spin-orbit interaction and averaged over spin branches takes the form
\begin{multline}
\label{Odd viscosity: eta_xy,SO}
\eta_{xy,SO} = \frac{\eta_{xy,SO}^+ + \eta_{xy,SO}^-}{2} =
P_s \varepsilon_F \frac{\partial\eta_{xy,SO}^s}{\partial \varepsilon_F},
\end{multline}
where $P_s \ll 1$ is the electron spin polarization degree 
\begin{equation}
P_s \equiv \frac{N^+ - N^-}{N^++N^-}= -\frac{g\mu_B B}{2\varepsilon_F},
\end{equation}
$g$ is the electron g-factor, $\mu_B$ is the Bohr magneton. As can be seen from these expressions, the spin-orbit contribution to the odd viscosity arises only for nonzero electron polarization, i.e., it requires breaking of time inversion symmetry. For electrons with a given spin projection, $\eta^\pm_{xy,SO} = - \eta^{\pm}_{xy,SO}\ne 0$, but these contributions obviously have opposite signs, see Eq.~\eqref{so:eta}. The odd viscosity \eqref{Odd viscosity: eta_xy,SO} depends on the type of impurity potential. In particular, as already noted above, for short-range scatterers $\eta_{xy,SO}=0$. To analyze the dependence of $\eta_{xy,SO}$ on the disorder correlation length, consider, as an example, Gaussian impurities
\begin{equation}\label{Odd viscosity: U_c,v gauss}
U_{c,v}(q) = U_0 \cdot e^{-q^2r_0^2/4},
\end{equation}
where $U_0 = \int U_{c,v}(\mathbf r) d^2r$ is the "strength" of the impurity potential, $r_0$ is the correlation length. The dependence of odd viscosity on the dimensionless correlation length $k_Fr_0$ is shown in Fig. \ref{fig:2 eta_gauss}. In the limit $k_F r_0 \gg 1$, we have
\begin{equation}\label{Odd viscosity: eta_xy,SO^gauss}
\eta_{xy,SO}^{gauss} = \frac{4P_s}{\sqrt{3}}\cdot \xi \cdot k_F^2r_0^2\cdot\frac{\varepsilon_F^2}{U_0\hbar n_{imp}}.
\end{equation}

The expression for odd viscosity \eqref{Odd viscosity: eta_xy,SO}, due to the presence of smooth disorder and related to spin-orbit interaction, is one of the main results of this work.

\textbf{4. Spin-dependent hydrodynamic equations.} The kinetic equation~\eqref{Model:general form of kinetic equation}, supplemented by appropriate boundary conditions at the system edges and the Poisson equation relating $\mathbf E_H$ and the electron distribution function, allows, in principle, investigation of magnetotransport effects, including the anomalous Hall effect, for any type of disorder. In the case of smooth disorder, $\tau_n$ rapidly decreases with increasing angular harmonic number $n$, hence in the expansion~\eqref{non:eq} of the nonequilibrium distribution function in angular harmonics, it is sufficient to retain contributions with $n=1$ and $2$. This significantly simplifies the problem and reduces the description of electronic transport to a hydrodynamic model. In particular, summing~\eqref{Model:general form of kinetic equation} over momenta, one can obtain continuity equations for $N_\pm$ in standard form. Equations for fluxes are conveniently presented as a system of equations for electric current $\mathbf j \equiv e N^+\mathbf u^+ + e N^-\mathbf u^-$ and for spin current (current of the $z$-component of spin) $\mathbf i^{z} = eN^+\mathbf u^+ - e N^-\mathbf u^-$. The factor $e$ in the expression for $\mathbf i^z$ is introduced for convenience. Moreover, we represent each of the currents as a sum of a ``normal'' component, which does not contain spin-orbit contributions and can be calculated at $\xi=0$, and an ``anomalous'' (spin-orbit) one, linear in $\xi$: $\mathbf j = \mathbf j_n + \mathbf j_a$, $\mathbf i^z=\mathbf i^z_n +\mathbf i^z_a$. Correspondingly, the Hall field $\mathbf E_H = \mathbf E_H^n+ \mathbf E_H^a$. Ultimately, we will be interested in $\mathbf j_a$, the spin-orbit contribution to the electric current, and the corresponding component of the Hall field $\mathbf E_H^a$. 
We emphasize that within the framework of our model, where electron-electron collisions are absent, the spin current relaxes only similarly to the electric current, and spin relaxation manifests itself only in higher orders in the parameter $\xi$ and can be neglected.

In particular, the ``normal'' component of electric current obeys the standard equation,
\begin{subequations}\label{anom:all}
\begin{multline}\label{Anomalous hydro:normal navier-stockes}
\frac{\mathbf j_n}{\tau_1} -\omega_c [\mathbf j_n \times \hat{\mathbf z}] - \frac{e^2N}{m}(\mathbf E + \mathbf E_H^n) = \\ = \eta\Delta \mathbf j_n + \eta_{xy,H}[\Delta \mathbf j_{n}\times \hat{\mathbf z}],
\end{multline}
containing the Lorentz force and odd viscosity related to the Lorentz force~\eqref{Hall:eta}. All parameters $\tau_1,\eta,\eta_{xy,H}$ are taken at energy equal to the Fermi energy; thus, the current depends on $\varepsilon_F$ as a parameter. Similarly, the ``normal'' component of spin current, taking into account possible coordinate dependence of the degree of spin polarization $P_s = P_s(\mathbf r)$, obeys the equation
\begin{multline}
\label{isn}
e\frac{\varepsilon_FN}{m}\mathbf \nabla P_s - P_s\frac{N}{m}e^2\mathbf E - \zeta\cdot [e^2\mathbf E\times[\mathbf \nabla \times \mathbf i_{n}^z]] \\
= \eta \Delta \mathbf i_n^z + \eta_{s}\Delta \mathbf j_n - \frac{\mathbf i_{n}^z}{\tau_1} - \frac{\mathbf j_n}{\tau_1^{s}},
\end{multline}
where the following notations are introduced: $\zeta = \eta/\varepsilon_F + \partial\eta/\partial\varepsilon_F$, $\eta_{s} = P_s\varepsilon_F\cdot \partial \eta /\partial \varepsilon_F$ and $(\tau_1^{s})^{-1} = P_s\varepsilon_F\cdot \partial\tau^{-1}_1/\partial \varepsilon_F$. Note that the flow of electric current is accompanied by the emergence of spin current both directly due to the presence of polarization in the system and due to the dependencies of momentum relaxation time and viscosity on Fermi energy, leading to their difference for electrons with different spin projections.

The equation for the "anomalous" component of electric current, taking into account the induced anomalous Hall field $\mathbf E_H^a$, has a somewhat more complex form:
\begin{multline}\label{ah:ans}
\frac{\mathbf j_a}{\tau_1}+ \frac{\mathbf \nabla p}{m} - \frac{N}{m}e^2\mathbf E_H^a \\= \eta \Delta \mathbf j_a + \eta_s \Delta\mathbf i^z_{a} + \eta_{xy,SO}[\Delta \mathbf j_n \times \hat{\mathbf z}] + \eta_{xy,SO}^s[\Delta \mathbf i_{n}^z\times \hat{\mathbf z}] \\
+ \Gamma_1\,[\mathbf j_n\times \hat{\mathbf z}] + \gamma_1\,[\mathbf i^z_{n}\times \hat{\mathbf z}] - \frac{\mathbf i_a^z}{\tau_1^s},
\end{multline}
where $\Gamma_1 = P_s\varepsilon_F\cdot \partial \gamma_1/\partial \varepsilon_F$ and
\begin{equation}
\frac{\mathbf \nabla p}{m} = P_s\left(\eta\,\Delta \mathbf i^z_{a} + \eta_{xy,SO}^s[\Delta \mathbf j_n \times \hat{\mathbf z}] + \gamma_1\, [\mathbf j_n\times \hat{\mathbf z}] - \frac{\mathbf i^z_a}{\tau_1}\right),
\end{equation}
is the "anomalous" pressure contribution from the zeroth harmonic of the distribution function, due to the fact that $N_+ - N_-$ may depend on coordinates. This contribution is important but leads only to renormalizations of the corresponding coefficients on the right side of the equation for anomalous current \eqref{ah:ans}. Finally, the equation for the anomalous part of spin current has the form
\begin{multline}
\label{isa}
e\frac{\varepsilon_FN}{m}\mathbf \nabla P_s - \omega_c[\mathbf i^z_{a}\times \hat{\mathbf z}] \\ = \eta\, \Delta \mathbf i_a^z + \eta_{xy,H}[\Delta \mathbf i^z_a\times \hat{\mathbf z}] + \eta_{xy,SO}^s[\Delta \mathbf j_n\times \hat{\mathbf z}] \\ + \gamma_1[\mathbf j_n \times \hat{\mathbf z}] - \frac{\mathbf i_a^z}{\tau_1} - \frac{\mathbf j_a}{\tau_1^s}.
\end{multline}
\end{subequations}
As above, all quantities depend on $\varepsilon_F$ as a parameter.
The system of equations for spin and electric currents \eqref{anom:all} is another main result of this work.

In an unbounded homogeneous system, electric and spin currents do not depend on coordinates. For $\mathbf B=0$, Eq.~\eqref{isa} describes the relationship between spin current and electric current due to asymmetric scattering $\mathbf i_a^z = \gamma_1 [\mathbf j_n \times \hat{\mathbf z}]$~\cite{2020arXiv200405091G}. If $\mathbf B\ne 0$, then Eq.~\eqref{Anomalous hydro:normal navier-stockes} describes the relationship between field and current due to the normal Hall effect, and from~\eqref{ah:ans} one can derive the anomalous Hall conductivity for two-dimensional electrons in the form
\begin{equation}
\sigma_{xy}^a = - \sigma_{yx}^a = P_s\frac{e^2N}{m}\frac{\partial}{\partial \varepsilon_F}(\gamma_1\tau_1^2\varepsilon_F),
\end{equation}
where $N=N^++N^-$.

\begin{figure}[tb]
\includegraphics[width=0.95\linewidth]{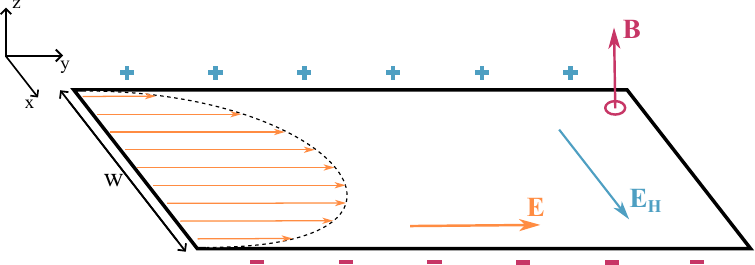}
\caption{Schematics of an electron channel, infinite along the $y$ axis and of width $w$ along the $x$ axis. The electric field is directed along the channel ($\mathbf E \parallel y$), the magnetic field is perpendicular to the plane of particle motion ($\mathbf B \parallel z$). The Hall effect leads to charge accumulation at the channel edges and $\mathbf E_H \parallel x$. The figure shows the parabolic current profile along the channel in the limit of hydrodynamic electron transport.}
\label{fig:2}
\end{figure}

\textbf{5. Anomalous Hall effect in a conducting channel.} As an example of applying the developed theory, consider a conducting channel of width $w$ along $x$, along which an external electric field $\mathbf E \parallel y$ is directed, see Fig.~\ref{fig:2}. The magnetic field $\mathbf B \parallel z$ is perpendicular to the plane of electron motion. In this case, a Hall field $\mathbf E_H \parallel x$ appears. In this geometry, there are no currents along the $x$ axis. Let us find the $y$-component of electric current in the quasi-hydrodynamic limit, where $l_1 \gg w \gg l_2$ and $l_{1,2} = v_F\tau_{1,2}$ are the relaxation lengths of momentum and the second angular harmonic of the distribution function. Solving \eqref{Anomalous hydro:normal navier-stockes} with boundary conditions of zero current at the channel walls, in limit $l_1l_2\gg w^2$, we obtain~\cite{glazov_zphrabyan:ftt, grigoryan2023anomalous, Glazov_2021b}
\begin{equation}\label{AHE:normal current}
\mathbf j_{n} = \frac{e^2N\mathbf E}{2m\eta}\left[\left(\frac{w}{2}\right)^2 - x^2\right],
\end{equation}
which corresponds to Poiseuille flow with a parabolic profile. Substituting \eqref{AHE:normal current} into \eqref{anom:all}, and taking into account boundary conditions and the continuity equation, we obtain the following expression for the anomalous contribution to the Hall field:
\begin{equation}\label{Hall effect: E_H^a}
\mathbf E_H^a = [\hat{\mathbf z} \times \mathbf E]\left\{\left( \frac{\Gamma_1}{\eta} - \gamma_1\frac{\eta_s}{\eta^2}\right)\left[\left(\frac{w}{2}\right)^2 - x^2\right]-\frac{\eta_{xy,SO}^H}{\eta}\right\},
\end{equation}
where the first contribution corresponds to the contribution of asymmetric scattering \cite{glazov_zphrabyan:ftt, grigoryan2023anomalous}, and the second contribution is from odd viscosity, written in the form in which Hall viscosity usually enters the expression for field $\mathbf E_H^n$ \cite{PhysRevB.106.245415}. With such notation, the ``effective'' odd viscosity will have the form $\eta_{xy,SO}^H = \eta \cdot P_s\varepsilon_F\, \partial \gamma_2\tau_2/\partial \varepsilon_F$, see the red curve in Fig. \ref{fig:2 eta_gauss}.
The first contribution is possible in any system and at any disorder potential, including short-range, and the second contribution, due to odd viscosity, is possible only in systems where the electron flow is non-uniform.
In the case of a short-range potential, the contribution of odd viscosity is absent. In the quasi-hydrodynamic regime, where $l_2 \ll w \ll l_1$, the contribution of odd viscosity due to impurity scattering can be significant.
Anomalous contributions to the Hall effect can be separated from normal one in electron paramagnetic resonance experiments by ``erasing'' the spin polarization with a high-frequency field, or in low-symmetry structures in a inclined magnetic field, where the Lorentz force and spin-orbit contributions can be controlled by different components of the external field.

It is worth to mention that, generally speaking, in the hydrodynamic regime, the boundary corrections to the Hall field can have the same order as the contribution from the odd viscosity~\cite{PhysRevB.106.245415}. In this regard, studying the anomalous Hall effect caused by electron polarization can help elucidate these phenomena.

\textbf{6. Conclusion.} The work presents a microscopic theory of odd (off-diagonal) viscosity of electrons in two-dimensional systems with smooth disorder and spin-orbit interaction. It is shown, that such contribution occurs even in absence of electron-electron collisions. Analytical expressions for the components of the viscosity tensor are obtained, including the contribution of spin-orbit interaction $\eta_{xy,SO}$, which is absent in the case of short-range scattering but significant for a long-range impurity potential. It is shown that this contribution is proportional to the spin polarization of carriers, the spin-orbit interaction constant, and depends on disorder parameters.

A closed system of equations for spin and electric currents is derived, taking into account both normal and anomalous contributions for the quasi-hydrodynamic transport regime, where the relaxation time of the first angular harmonic of the distribution function significantly exceeds the relaxation times of higher harmonics. The contribution of odd viscosity to the anomalous Hall field is obtained. This work is the first step towards constructing a microscopic theory of odd viscosity in interacting electronic systems.

\textbf{Acknowledgments.}
The work was supported by the RSF Project No. 22-12-00211-Continuation.

\textbf{Conflict of Interest.} The authors of this work declare that they have no conflict of interest.

\bibliographystyle{ieeetr}
\bibliography{bibexport}

\end{document}